\documentclass[twocolumn]{bioRxiv}
\usepackage{capt-of}

\usepackage{nameref}
\usepackage{float}
\usepackage{setspace}
\usepackage{svg}
\usepackage{csquotes}

\AtBeginBibliography{\scriptsize}

\begin{document}
\leadauthor{McKenzie-Smith}
\title{Capturing continuous, long timescale behavioral changes in \textit{Drosophila melanogaster} postural data}
\shorttitle{Capturing continuous, long timescale behavioral changes in \textit{Drosophila}}

\author[1,\Yinyang]{Grace C. McKenzie-Smith \orcidlink {0000-0001-8428-0043}}
\author[2,\Yinyang]{Scott W. Wolf \orcidlink{0000-0003-4397-1395}}
\author[2,3]{Julien F. Ayroles \orcidlink{0000-0001-8729-0511}}
\author[1,2,\href{mailto:shaevitz@princeton.edu}{\Letter}]{Joshua W. Shaevitz \orcidlink {0000-0001-8809-4723}}
\affil[1]{Department of Physics, Princeton University, Princeton, NJ, USA}
\affil[2]{Lewis-Sigler Institute for Integrative Genomics, Princeton University, Princeton, NJ, USA}
\affil[3]{Department of Ecology and Evolutionary Biology, Princeton University, Princeton, NJ, USA}
\affil[\Yinyang]{These authors contributed equally}
\date{}

\maketitle
\begin{abstract}
Animal behavior spans many timescales, from short, seconds-scale actions to circadian rhythms over many hours to life-long changes during aging. Most quantitative behavior studies have focused on short-timescale behaviors such as locomotion and grooming. Analysis of these data suggests there exists a hierarchy of timescales; however, the limited duration of these experiments prevents the investigation of the full temporal structure. To access longer timescales of behavior, we continuously recorded individual \textit{Drosophila melanogaster} at 100 frames per second for up to 7 days at a time in featureless arenas on sucrose-agarose media. We use the deep learning framework SLEAP to produce a full-body postural data set for 47 individuals resulting in nearly 2 billion pose instances. We identify stereotyped behaviors such as grooming, proboscis extension, and locomotion and use the resulting ethograms to explore how the flies' behavior varies across time of day and days in the experiment. We find distinct circadian patterns in all of our stereotyped behavior and also see changes in behavior over the course of the experiment as the flies weaken and die.
\end{abstract}


\begin{corrauthor} 
    Joshua W. Shaevitz <\href{mailto:shaevitz@princeton.edu}{shaevitz@princeton.edu}>
\end{corrauthor}

\section*{Significance Statement}
Animal behaviors exist on many timescales, ranging from the milliseconds required for speaking individual words to the years of behavioral shifts due to aging. Investigating the temporal structure of behaviors at longer timescales is challenging, and requires continuous, high resolution data taken over days. Here we present a data set of continuously captured high resolution \textit{Drosophila melanogaster} behavior recorded over 4-7 days. Our continuous, high resolution data allows us to describe patterns in fine-grained behaviors such as locomotion speed, proboscis extension, and grooming across minutes, hours, and days. With this data we reveal detailed circadian cycles of behavior and trends of behavior over the lifetime of the fly.
\section*{Introduction}

\begin{figure*}[!ht]
	\centering
	\includegraphics[width=\linewidth]{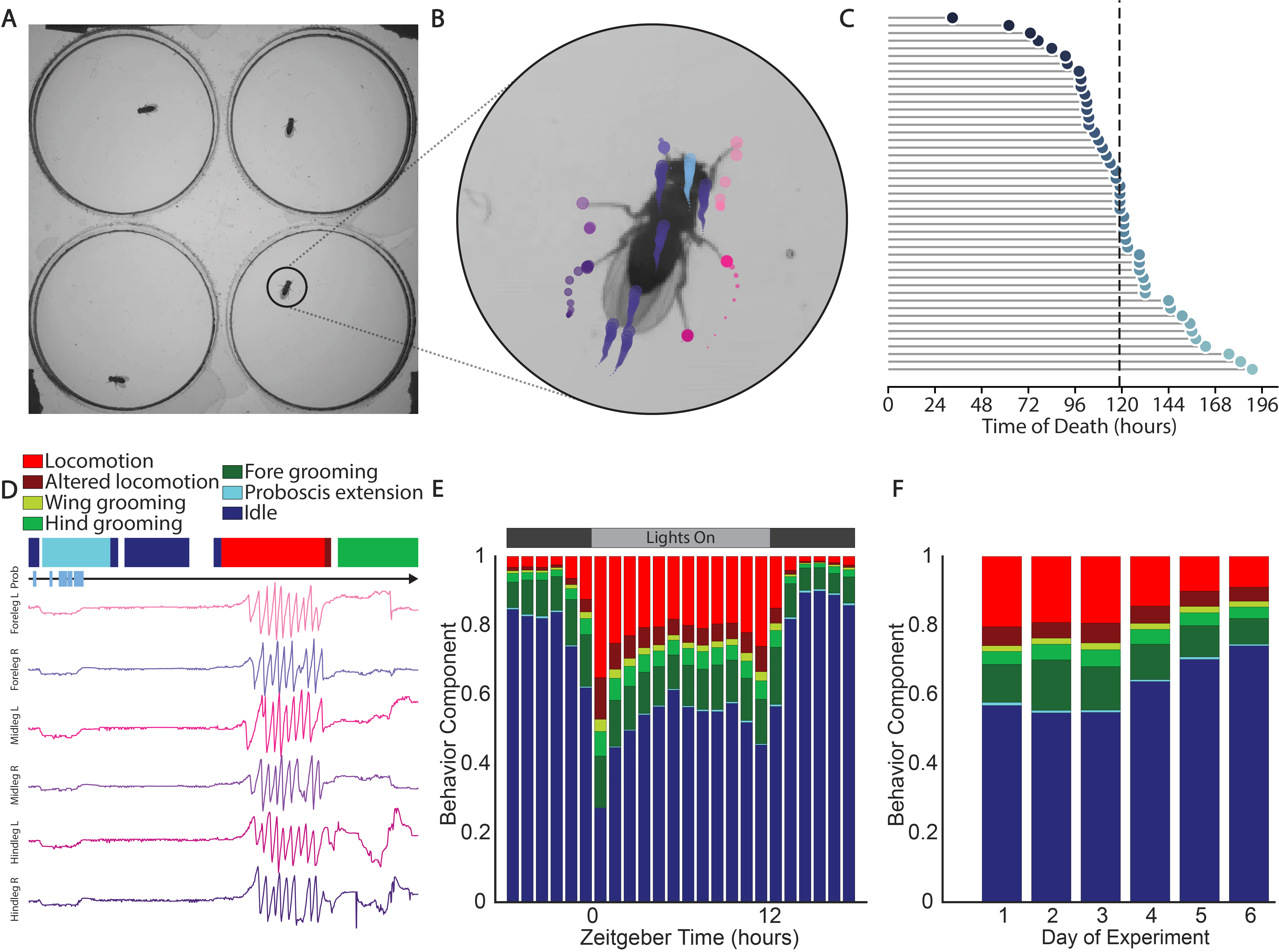}
	\caption[Experimental schematic showing tracking, lifespans, and behavioral segmentations across timescales.]{Experimental schematic showing tracking, lifespans, and behavioral segmentations across timescales. \textbf{A} Image showing the experimental arena as viewed from below. The behavior of 4 \textit{D. melanogaster} is captured simultaneously while giving each fly enough room to freely carry out all behaviors except flight. \textbf{B} Magnified view of a single individual showing tracks for each node of the SLEAP skeleton. Each color denotes a node and circle sizes increase with time. \textbf{C} Survival curve of the 47 flies included in the experiment. Death occurs on average after $\sim$119 hours, or almost 5 full days into the experiment. \textbf{D} Ethogram and egocentrized traces for each tarsi and a raster denoting proboscis visibility. \textbf{E} Barplot showing the geometric means of stereotyped behavior components across all flies and all complete 24h periods. \textbf{F} Barplot showing the geometric means of stereotyped behavior components across all flies and all hours grouped by experimental day.}
	\label{fig:figure1}
\end{figure*}

Uncovering the temporal structure of behavior has long been a topic of theoretical interest and experimental challenge \autocite{Dawkins1976-xr,Brown2018-qj,Berman2018-vy,Bialek2022-le}. Animals carry out sequences of behaviors on many timescales, from the short timescales of the individual movements required for grooming, eating, and social communication to the longer timescales of hunger, arousal, circadian cycles, mating seasons, and the aging process. The specifics of these behavior sequences determine much of what we can characterize about an animal, such as its health, reproductive fitness, and that idiosyncrasy of action that we might call `personality.' These behavior sequences also give us indirect ways to assess the internal processes of the animal, such as neural activity, gene expression, and other internal states like hunger or fatigue. Finding general principles that govern the order of behaviors would be an exciting step forward in understanding how animals interact with the world around them and how internal factors may shape that interaction. This course of study requires data that covers the many timescales over which animal behavior varies.

Historically, taking long-timescale data covering days or weeks of an animal's life has required balancing continuity, throughput, and dimensionality. In \textit{Drosophila melanogaster}, simple experimental setups, such as beam-break assays, allow for continuous monitoring of activity levels over days \autocite{Pfeiffenberger2010-qu,Harbison2017-cs}, but fail to capture the high-resolution data necessary for modern techniques of behavior analysis such as MotionMapper \autocite{Berman2014-ee}, B-SOiD \autocite{hsu2021b}, VAME \autocite{luxem2022identifying}, or Keypoint-MoSeq \autocite{weinreb2023keypoint}. On the other hand, the acquisition of high-resolution data has been restricted to short timescales by the computational resources required to store and process the extremely large imaging data, imposing an upper limit on the order of an hour. When studying fine-grained behavioral variation at longer timescales, previous work utilized short recordings taken from different individuals with ages distributed across the lifespan of the animal \autocite{Overman2022-gk}.

Here, we leverage recent computational advances to record a high-resolution continuous data set of \textit{D. melanogaster} behavior spanning 4-8 days. We recorded 47 freely moving \textit{D. melanogaster} using constant IR illumination and an IR-sensitive camera at a frame rate of 100Hz, with a 12-hour visible-light day/night cycle. We tracked 14 body parts from each fly using SLEAP \autocite{Pereira2022-as} and utilized MotionMapper to characterize stereotyped behavioral states, such as grooming, locomoting, and feeding. Using techniques of compositional data analysis \autocite{Aitchison_1986}, we characterize the dynamics of this behavioral repertoire across time of day and over the days of the experiment. We find distinct circadian patterns in all measured behaviors, including grooming, proboscis extension, and locomotion speed. We see an overall decline in circadianicity, the difference in behavior between day and night hours, across days in the experiment as flies weaken an die, and see general declines in feeding and locomotion speed as the fraction of time spent in an idle state increases. Overall, we find that our data captures both expected and novel patterns of \textit{D. melanogaster} behavior across multiple 24-hour periods. We also provide this data to the broader community as a resource to study \textit{D. melongaster} behavior as it evolves along timescales beyond the scope of previous research.

\section*{Results and discussion}
% Here for placement on prev page
\begin{figure*}[!ht]
\centering
\includegraphics[width=\linewidth]{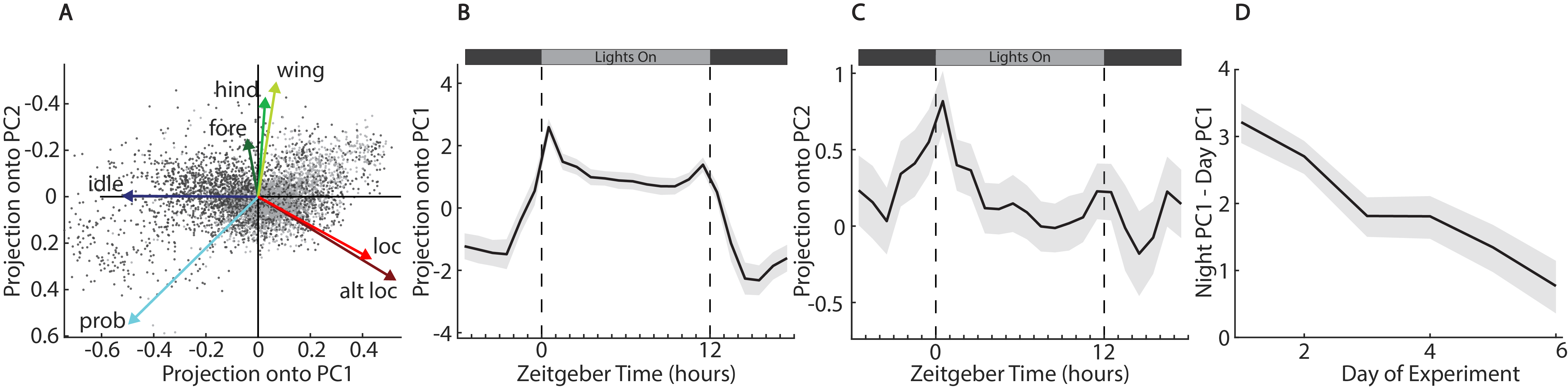}
\caption[Principal component analysis of stereotyped behavioral components of all flies across all experimental hours.]{Principal component analysis of stereotyped behavioral components of all flies across all experimental hours. \textbf{A} Biplot showing the projections of individual fly hours and loadings of each stereotyped behavioral component. Dark gray dots show timepoints from when lights are off and light gray dots show timepoints when lights are on.  \textbf{B} Projection of PC1 against time of day for all complete 24h periods of all flies. \textbf{C} Projection of PC2 against time of day for all complete 24h periods of all flies. \textbf{D} Circadianicity vs day of experiment as measured by the difference between the average projection onto PC1 during night hours minus the projection onto PC1 during day hours.}
\label{fig:figure2}
\end{figure*}

We designed a recording apparatus to allow for continuous capture of \textit{D. melanogaster} behavior over the course of days (see \hyperref[sec:methods]{Methods} for details and \autoref{fig:figures1}A). \textit{D. melanogaster} were constantly illuminated from above with IR light, to which they have minimal visual sensitivity \autocite{De_Salomon1983-ip}, while LED panels provided a 12-hour visible-light day/night cycle with the same on/off times under which the animals were raised. We made arenas by layering pieces of transparent laser-cut acrylic to create cylindrical chambers in which flies lived and behaved over the course of our experiments (\autoref{fig:figures1}B). We limited the arenas to 1.5mm in height to prevent flying and to decrease the incidence of wall walking and ceiling walking, which lower tracking quality. We provided the flies with a base gel layer of sucrose-agarose, which permitted survival of up to 7 days while preventing the significant fungal growth observed when yeast extract was included. We recorded four freely behaving \textit{D. melanogaster} in individual chambers per camera at 100 Hz with a resolution of 28.25 pixels/mm (\autoref{fig:figure1}A). This is sufficient to resolve relevant features of the \textit{D. melanogaster} body such as the tarsi (leg tips) and proboscis (\autoref{fig:figure1}B).

For each experiment, we imaged male \(\textit{iso}^{KH11}\) \textit{D. melanogaster} from two days post-eclosion (emergence from the pupa as an adult insect) until death, yielding 4-8 days of continuous recording with half the flies dying by Day 5 (\autoref{fig:figure1}C), for a total of 5,584 fly-hours. Note that this lifespan is shorter than conventional assays due to the nutrient-limited sucrose-based food source we used to avoid fungal growth \autocite{hassett1948utilization}. 

 In the natural world, daytime conditions increase lighting and temperature, but in the lab, the \textit{D. melanogaster} circadian cycle can be maintained by these factors in isolation, with a day/night lighting cycle under constant temperature or a temperature cycle under constant lighting conditions \autocite{glaser2005temperature,zhang2010light,vanin2012unexpected}. Our experiments have both a change in light intensity and temperature conditions between day and night, with daytime temperature levels varying between different experiments ($\sim$28-29 $^{\circ}$C for experiments 1-2, and $\sim$ 30-31 $^{\circ}$C for experiments 3-4) and nighttime temperatures settling to $\sim$27 $^{\circ}$C for all experiments (see \hyperref[sec:methods]{Methods} and \autoref{fig:figures9}). We provide temperature and humidity recordings with our dataset.

To extract postural information from our data, we used SLEAP, a deep-learning-based framework that can infer animal posture based on user-generated training data \autocite{Pereira2022-as} (See \hyperref[sec:methods]{Methods} for details). We tracked a 14-point skeleton comprised of the head, eyes, thorax, abdomen, wing tips, tarsi, and proboscis of each individual (\autoref{fig:figure1}B). While our mean localization error was less than .1mm (\autoref{fig:figures2}), the quality of the tracks decreased when animals walked along the edges of the arenas. Accordingly, we built a classifier to identify the time points when flies walked on the edges (\autoref{fig:figures3}), and excluded these time points from the portions of our analysis reliant on accurate tracking of any body part but the thorax.

In order to quantify discrete, stereotyped behaviors, we modified the MotionMapper pipeline \autocite{Berman2014-ee} to parallelize more steps and optimize use for postural data instead of raw images (\autoref{fig:figures4}). We used the Lomb-Scargle periodogram rather than a continuous Morlet wavelet transform to generate power spectra for each body part as this algorithm does not require interpolation of missing data. As a first step, we classified all time points with a total power of less than $0.5012mm^2$, summed over all tracked positions, as `idle', i.e. times where the flies are not moving at all. We exclude all time points classified as idle and all non-idle edge time points from the spectral analysis that follows. `Idle' and `non-idle edge' then become their own behavioral categories. 

Our total amount of data is too large to allow for direct classification of behaviors from all time points. Instead, we generated a set of 141 one-hour videos sampled evenly across flies, time of day, and day of experiment. From this subset of the videos, we selected 64,014 time points representative of the full suite of observed dynamics via an importance-sampling algoirthm \autocite{Berman2014-ee}, and embed the power spectra from these points into two dimensions using the UMAP algorithm.

We then embedded all time points from the 141 hour subset and found well-separated peaks of high density using an adaptive threshold (\autoref{fig:figures5}). We assigned behavior labels to these regions by looking at randomly selected clips from time points where the flies' dynamics fell within a given region's boundaries for a reasonable length of time. We grouped together regions of similar dynamics, and identified seven well-defined behaviors: idle, proboscis extension, fore grooming (of the eyes or forelegs), hind-grooming (of the abdomen or hindlegs), wing grooming, altered locomotion (often involving slipping or limping), and locomotion. The idle behavior state includes all points assigned as idle using the total power cutoff as well as several regions of the spectral embedding that contained idle behaviors with single-limb tracking errors. In addition to these well defined behaviors, $\sim$15\% of all time points represent unstereotyped dynamics, where the fly is either on the edge and non-idle, or where its dynamics fall outside the boundaries of the identified peaks of stereotyped behaviors. We exclude these time points from later analyses. Finally, we project the full data set into the  two-dimensional space and use the behavioral boundaries from the training set to classify each time point as one of the six stereotyped behaviors, idle, non-idle edge, or unstereotyped.

We used 5 frames (1/20 of a second) as a minimum bout length for each stereotyped behavior, and forward-filled each fly's behavior sequence with this bout length, assigning any bout of 4 frames or less to the previous region of long duration. The resulting ethograms permit analysis of patterns in locomotion, feeding, and grooming (\autoref{fig:figure1}D). Because our data is continuous over multiple 24-hour periods, we can look at how behavior varies with time of day and across days of the experiment. 

Our data is closed (i.e. the fraction of time spent in all behaviors must add up to one) requiring us to use methods of compositional data analysis to examine changes across flies, hours, and days \autocite{Aitchison_1986,matcomp}. Averages of closed data are best calculated as geometric means, which we denote `behavior components'. To discuss circadian behavioral effects, we use Zeitgeber time (ZT), where time is measured from the onset of a periodic stimulus rather than from midnight on a clock, to capture the cyclic nature of circadian effects. For this set of experiments, ZT = 0h corresponds to the visible lights coming on, and ZT = 12h corresponds to lights turning off. 

Looking across all fly hours and all days, we see a distinct circadian pattern of behavior with higher levels of idle during the dark hours, and more locomotion and grooming during the light hours (\autoref{fig:figure1}E). The first hour after the lights turn on is particularly distinct, with comparatively high levels of locomotion and grooming. The locomotion and grooming behavior components increase in the hours leading up to lights on and lights off, indicating anticipation of the change in lighting condition. Over the course of the experiment, the flies' behaviors start changing significantly after Day 3 (\autoref{fig:figure1}F). Time spent in idle increases over Days 4 through 6 as flies begin dying on the nutritionally incomplete food used for this experiment.

To examine overall behavior variation across hours of the day, we carried out a principal components analysis of the compositional data \autocite{aitchison2002biplots,filzmoser2009principal} using the compositions package in R \autocite{van_den_boogaart_compositions_2008}. We used the isometric log-ratio transformed behavior compositions to carry out robust principal components analysis using the Minimum Covariance Determinant (MCD) method, and then backtransformed the result into centered log-ratio loadings. The first three principal components (PCs) explain $\sim$85\% of the variance across all fly hours (\autoref{fig:figures6}). 

As can be seen in the biplot of the first two PCs (\autoref{fig:figure2}A), PC1 largely weights the locomotion behaviors, locomotion and altered locomotion, against idle and proboscis extension. This PC describes the main differences between day and night, with positive projection averages during the day, corresponding to more locomotion/grooming, and negative values at night when the animals are idle (\autoref{fig:figure2}B). The average projection along PC1 begins to increase before the lights turn on, indicating that the animals anticipate the rise of the sun. Peak amplitude along this PC occurs just after the lights turn on, potentially indicating a slow morning transition from nighttime behaviors to daytime activity. The level of this projection stays roughly constant throughout the day, but then increases and peaks just before the lights turn off at 12h ZT. This is followed by a slow decline in the amplitude after dark until reaching a steady night level.

Previous behavioral studies of the \textit{D. melanogaster} circadian cycle have used relatively coarse metrics, such as the activity counts generated by \textit{Drosophila} Activity Monitors \autocite{Pfeiffenberger2010-qu}. These studies have shown that \textit{D. melanogaster} have peaks of locomotion activity around their subjective morning and evening, with the increase in activity slightly anticipating the actual change in lighting conditions \autocite{Chiu2010-qa,Dubowy2017-bg}. Our high-resolution behavioral data and the projection along PC1 recapitulate these general trends, but show quantitative difference when the lights change. In particular, we see gradual change in amplitude after lights turn off that last several hours whereas this previous work sees a more abrupt sesation of locomotion at this time.

The second principle component weights the three grooming behaviors (fore, hind, and wing) against the locomotion behaviors and proboscis extension. The average projection onto PC2 has a distinct peak during the hour just after lights turn on, separating this unique part of the circadian pattern from the more general day vs. night changes in behavior picked up by PC1 (\autoref{fig:figure2}C). PC3 largely separates the first two experiments (begun 02/17/2022 and 03/13/2022) from the second set of experiments (begun 03/26/2022 and 04/18/2022) (\autoref{fig:figures7}). The second set of experiments took place at higher temperatures (\autoref{fig:figures9}). The difference in the projections of each fly-hour along PC3 between the two sets of experiments is lowest on Day 1, and increases over experimental days.

Since the amplitude along PC1 largely follows the day-night cycle and describes the circadian change in behaviors, we use the difference between the average value of PC1 during light and dark hours to define a `circadianicity' value for each fly day. We find that circadianicity decreases steadily over days in the experiment (\autoref{fig:figure2}D). Previous studies have found that the sleep/wake cycles of behavior in \textit{D. melanogaster} weaken as they age \autocite{koh2006drosophila}. While the flies in our experiment were all comparatively young (all died before 10 full days, whereas life expectancy is 2-3 months under ideal conditions), they were living in very harsh conditions of relatively high temperature, low humidity, and poor nutrient availability. The gradual weakening over the course of the experiment is in some ways similar to an accelerated aging, and the steady decline in circadianicity over 6 days is similar to the decline in the strength of the sleep/wake cycle seen in over experiments over 60 days \autocite{koh2006drosophila}.

\begin{figure*}[!ht]
\centering
\includegraphics[width=\linewidth]{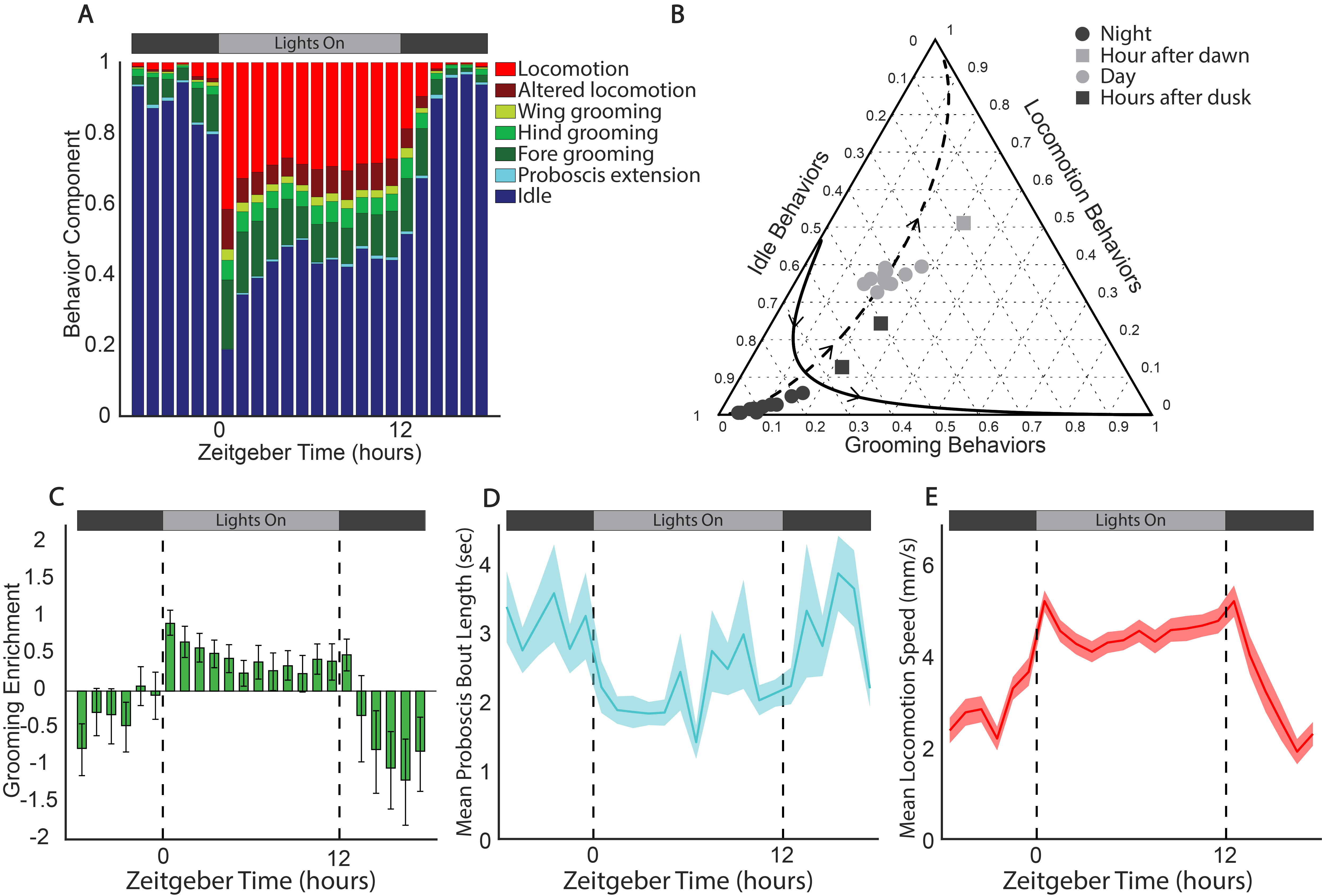}
\caption[Circadian patterns of behavior on experimental Day 1]{Circadian patterns of behavior on experimental Day 1. \textbf{A} Barplot showing the geometric means of stereotyped behavioral components of the first experimental day across all flies. \textbf{B} Ternary plot showing the geometric means of condensed behavioral components across all flies for each circadian hour of Day 1. Directions along PC1 (dashed) and PC2 (solid) as calculated by perturbing the geometric mean of the displayed data points \autocite{billheimer2001statistical}. The ternary plot was generated using the Ternary Plots package in MATLAB \autocite{mattern}. \textbf{C} Grooming enrichment with respect to the geometric mean of the condensed grooming behavioral component of the first experimental day for each fly with bootstrapped confidence intervals. \textbf{D} Mean proboscis bout length by hour of the first experimental day. The shaded region is the standard error. \textbf{E} Mean locomotion speed (mm/s) during stereotyped locomotion state by hour of the first experimental day. The shaded region is the standard error.}
\label{fig:figure3}
\end{figure*}

We leveraged our high-resolution behavior data to carry out an in-depth analysis of \textit{D. melanogaster} circadian patterns of behavior, focusing our analysis on the first day of the experiment when circadianicity was strongest(\autoref{fig:figure2}D). Flies were reared from embryos to two day old adults with the same light/dark cycle time and phase as the experiments. Even before eclosing, \textit{D. melanogaster} exhibit circadian patterns of certain behaviors, such as larval negative phototaxis \autocite{mazzoni2005circadian} or eclosion \autocite{pittendrigh1954temperature}, so it is unsurprising that even 2-3 day old adults already have a strong circadian pattern.

The geometric means of behavior components across all flies versus ZT for Day 1 show the expected pattern of increased idle during the night and increased locomotion during the day (\autoref{fig:figure3}A). The hour just after lights on remains the most distinct, with a very low idle behavior component. After lights off, the flies take $\sim$2 hours to settle into their characteristic high idle, low locomotion night state.

The temporal changes we observe in the two locomotion behaviors (altered locomotion and locomotion) are similar, as are the changes in the different grooming behaviors (fore, hind, and wing grooming). Using these strong correlations, we condensed our seven stereotyped behavior components into three categories, grouping together the grooming behaviors, the locomotion behaviors, and idle and proboscis extension. This allowed us to plot the average behavior composition for each circadian hour in a ternary plot to visualize differences in overall behavior across circadian time and along the previously identified PCs (\autoref{fig:figure3}B). The day and night hours cluster together and largely lie along PC1 as expected. The two hours just after lights off fall between these clusters as the flies transition into their night state of behavior. The hour just after lights on is an outlier, falling well off the line of variance explained by PC1, with higher proportions of grooming and locomotion behaviors compared to all other circadian hours. This hour lies in the direction of increasing PC2, which explains $\sim$17\% of the variance in the data. This, combined with the peak in the projection of behavior components along PC2 during the hour after lights on (\autoref{fig:figure2}C) indicates that this hour is a unique time point in the circadian cycle of behavior.

To further investigate the circadian pattern of grooming, we looked at the enrichment of grooming behaviors at each circadian hour compared to the geometric mean of the grooming behavior component across all hours (\autoref{fig:figure3}C). It has been shown that spontaneous grooming is under circadian control, but no clear pattern of when grooming happens throughout the day has been identified \autocite{qiao2018automated}. We find that grooming behaviors peak in the hour after lights on, contributing to the uniqueness of that time point, in agreement with our analysis of PC2. This temporary spike in grooming behavior may come from a need to refresh the various sensory appendages that lie along the body after a prolonged nighttime period without grooming.

Grooming remains enriched during the day, although this enrichment decreases after the early morning hours. Of the specifically identified grooming states, flies spend the most time grooming their fore limbs and eyes, with a lower proportion of time spent in hind grooming and wing grooming. This follows the flies' hierarchy of grooming motor programs, where fore grooming is prioritized, followed by abdomen grooming, which is captured in our hind grooming state, and finally wing grooming \autocite{seeds2014suppression}.

We also looked at daily eating patterns, using proboscis visibility as a proxy for feeding as proboscis extension is well correlated with food intake \autocite{wong2009quantification}. Previous studies report peak feeding activity centered around lights on and lights off in the mornings and evenings, with more feeding concentrated in the evening \autocite{Xu2008-pe,ro2014flic}. Proboscis extension comprises a very small fraction of our data, less than 1\% of the overall behaviors across all time points. Because it is such a small component, using compositional data analysis techniques is challenging, as many true zeros exist in the proboscis data. To get a better sense of the circadian nature of proboscis extension (and feeding), we instead look at the average duration of proboscis extension bouts over the course of the day (\autoref{fig:figure3}D). We find that flies typically leave their proboscis extended for about three seconds during night bouts, and about 2 seconds during day bouts. By this measure we do not see notable peaks of morning and evening feeding, but instead a more general trend of more time spent feeding at night, and less during the day.

Our observed trend of the locomotion behavior component with the time of day differs from results from previous studies using activity counts to measure overall movement levels. While there is a slight increase in the locomotion behavior component in anticipation of lights on in our study, it is less dramatic than the increase in activity counts observed in previous work \autocite{Chiu2010-qa}, and we see no peak in locomotion around lights off compared to the locomotion behavior component throughout the day hours. However, the circadian pattern of locomotion speed (the mean speed of flies only when they are in the `locomotion' state, calculated with the mean on a 5 frame rolling window) has peaks around each change in lighting conditions, along with anticipatory increases, particularly for lights on (\autoref{fig:figure3}E). In \textit{Drosophila} Activity Monitors, activity counts are recorded each time a fly crosses an infrared beam \autocite{Pfeiffenberger2010-qu}. These counts could increase due to a combination of increased movement time and increased movement speed. Our results indicate that it is an increase in movement speed, rather than time spent moving, that is responsible for the larger activity count peaks around lights on and lights off. The increase in locomotion speed before lights on, and the gradual falling off after lights off, indicates that flies are modulating their movement speed partially due to internal cues, rather than only as a startle response or some other reaction to lights on.

In addition to circadian patterns of behavior, the flies' behavior changed across experimental days as they weakened and died. Because of the nutritionally incomplete food and the relatively high temperature and low humidity, flies in our experiment all died within 8 days. The behavioral composition remained relatively constant across the first 3 days of the experiment, but starting at Day 4 the idle component began to increase (\autoref{fig:figure1}F). This is similar to the increase in the proportion of time male flies spend idle near the end of their lives in a more natural aging paradigm \autocite{Overman2022-gk}. Flies also show a reduction in the propensity to spend more time near the edge of the arena rather than the center after the first 3 days of the experiment (\autoref{fig:figures8}A). The wall following behavior of \textit{D. melanogaster} likely arises from boundary exploration, possibly as a means of seeking escape from a given enclosure \autocite{Soibam2012-wq}. Over the course of the experiment, flies decrease wall following behavior as habituation to an unchanging environment decreases exploration activities \autocite{Liu2007-gt}. However, the edge preference is difficult to disentangle from the differences in the fraction of time spent in other stereotyped behaviors at different radii (\autoref{fig:figures8}B). Flies spend an increased fraction of time in locomotion near the edge of the area and a increased fraction of idle near the center of the arena, and these differences may drive the observed changes in edge preference.

\begin{figure*}[!ht]
\centering
\includegraphics[width=\linewidth]{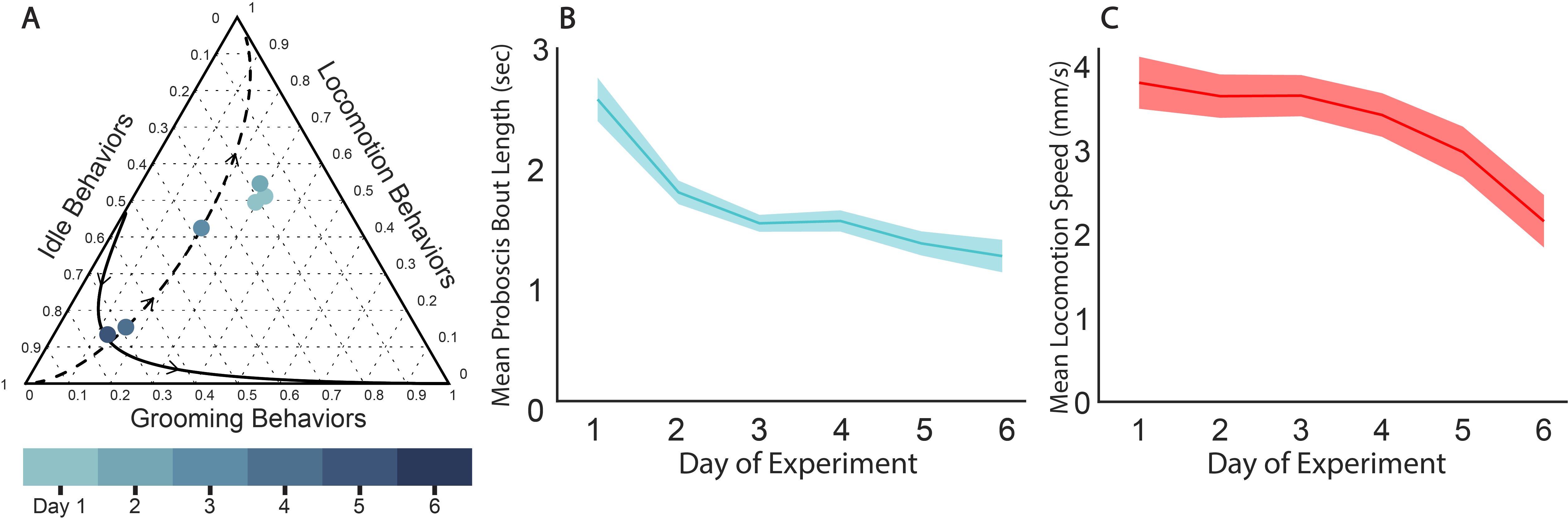}
\caption[Day wise behavioral changes throughout the experiment.]{Day wise behavioral changes throughout the experiment. \textbf{A} Ternary plot showing the geometric mean of the condensed stereotyped behavioral components of the first hour after lights on across all surviving flies for each complete 24h period. Directions of PC1 (dashed) and PC2 (solid) are also shown, as calculated based on perturbation of the geometric means of all circadian hours from Day 1. \textbf{B} Mean proboscis bout length by day of experiment. The shaded region is the standard error. \textbf{C} Mean locomotion speed (mm/s) during the stereotyped locomotion state by day of experiment. The shaded region is the standard error.}
\label{fig:figure4}
\end{figure*}

Since the hour after the lights turn on is such a unique time point in the circadian pattern of behavior, we were curious how the behavior components during that hour change over the days of the experiment. The geometric means of the relative proportions of the grooming behaviors, idle behaviors, and locomotion behaviors across all surviving flies in the hour after dawn remain similar for the first 3 days of the experiment, lying in a cluster offset from PC1 in the ternary plot (\autoref{fig:figure4}A). Starting at Day 4, however, the hour after dawn components begin falling onto PC1, and are much more similar to other circadian time points. This behavioral composition moves towards lower values of PC1 with age and becomes more similar to the nighttime composition. Thus, as the flies in our experiment weaken and die, not only do their day and night behavior patterns begin to look more similar, they also lose the distinctive behavioral character of the hour after dawn.

We also asked how feeding and locomotion change with age in our experiments. We find that proboscis bout duration decreased steadily through Day 3 and then plateaued (\autoref{fig:figure4}B). It has been reported that flies eat more as they age \autocite{driver1980metabolic}, but the limited food source and harsh environmental conditions may change this trend for the flies in our experiment. In contrast, the average locomotion speed remained steady through Day 3 and then began decreasing with age (\autoref{fig:figure4}C). The combination of steady locomotion speed and no increase in the fraction of time spent locomoting means that overall `locomotion activity', comparable to traditional activity counts, does not appreciably change over the first 3 days of the experiment after which there is a decline. Previous studies have shown that male flies in a natural aging context have an increase in locomotion activity during early life, before a decrease leading up to death \autocite{Le_Bourg1987-bx,Overman2022-gk}. We do not see this increase at young age in our experiment, however, lifelong locomotion patterns are genotype-dependent \autocite{fernandez1999differences}, so results from the\(\textit{iso}^{KH11}\) flies used here may not be not directly comparable to these previous studies.

\section*{Conclusion}

We report the first measurements of high resolution \textit{D. melanogaster} behavior recorded over many days with high temporal bandwidth. By leveraging recent advances in GPU-based video processing and postural inference, we captured the behavior of freely moving \textit{D. melanogaster} over the course of multiple days, encompassing the behavioral effects of circadian rhythms, starvation, aging, and habituation at continuous high resolution. Our data recapitulates many previously described trends in \textit{D. melanogaster} circadian and aging/dying patterns of behavior. We also leveraged high resolution postural data in combination with fine-grained ethograms to characterize changes in proboscis extension bout duration and locomotion speed across time of day and over the days of our experiment. With compositional data analysis techniques, we identified the hour after lights on as a uniquely distinctive time point in the circadian pattern of behavior.

Our data addresses several limitations of the high-quality ethological data currently available. Previous work on the temporal structure of behavior has found correlations extending beyond the length of the available data, typically 30-60 minutes \autocite{Berman2016-gk,Bialek2023-na}. The data presented here extends these time scales by more than two orders of magnitude. This data set is also the first to continuously capture high dimensional, high-resolution behavioral data across a circadian cycle, allowing us to investigate how changes in internal state related to time of day affect behavior. By recording when flies feed (as measured by proboscis extension), this data may also provide new insights into the effects of hunger and satiety. We provide both high-resolution recordings and our postural tracking output to facilitate further data analysis. The analyses presented here leverage only a fraction of the resolution and dimensionality provided by our data, and we hope this 100-fold increase in the amount of high-quality ethological data available will give rise to yet more tools and techniques. Finally, aging in our experiments was significantly accelerated due to nutrient limitation. Future work with new kinds of arenas and food sources may extend this type of high-resolution behavioral recordings to cover the full natural lifespan of a fly.

\phantomsection
\section*{Methods}
\label{sec:methods}
\footnotesize

\subsection*{Fly rearing}
To control for possible genetic effects, we used the highly inbred wild-type \(\textit{iso}^{KH11}\) strain. \(\textit{iso}^{KH11}\) flies were raised on standard cornmeal media (see \hyperlink{https://github.com/shaevitz-lab/long-timescale-analysis}{github.com/shaevitz-lab/long-timescale-analysis} for complete recipe) at $25^{\circ}$C under humidity 60\% with a 12-hour light/dark cycle, with visible light of $\sim$1300lux. Before each experiment, we performed egg lays and, on eclosion, flipped flies into new vials. We allowed the flies to age for two days, yielding 2-3 day-old flies, which we anesthetized using $\text{CO}_2$ and distributed males to arenas to be imaged.

\subsection*{Media}
During experiments, flies were allowed to feed ad lib from a pad of optically clear media (10\% sucrose, 1.5\% agarose). We were not able to include a protein source, such as yeast extract, as this led to high levels of fungal growth within 1-2 days that obscured imaging.

\subsection*{Arena}

We constructed experimental arenas out of laser-cut acrylic using acrylic cement (McMaster 7517A4) to adhere layers together (\autoref{fig:figures1}B). The bottom layer of each arena consisted of a 3mm layer of food (described above). Each individual fly was able to freely move about within a 25mm diameter cylinder of height 1.5mm. Because these arenas have straight walls, flies are able to walk along the sides, which can cause limb occlusions that pose difficulties to downstream postural tracking. To address this, we used a low arena height that impedes flies from easily maneuvering off the base layer. We also coated the top and walls with Sigmacote (Sigma-Aldrich SL2), which discourages flies from walking on the ceiling of the arena but does not fully restrict them from walking on the edges of the arenas.

\subsection*{Imaging and illumination}

The arenas are lit from above using 880nm IR LED pads (Advanced Illumination BL040801-880-IC). Below each arena, we placed high-resolution, high frame-rate cameras (FLIR BFS-U3-32S4M-C) paired with 880nm band-pass filters (ThorLabs FB880-70) (\autoref{fig:figures1}A). This combination allows bright, uniform lighting across the arenas permitting extremely short exposure times to reduce motion blur. Imaging from above and recording from below also eliminates condensation in the arenas. We found that the ideal balance between contrast and motion blur was at 1 ms exposure time. In addition, we used a pair of visible light LED panels at the top of the tent enclosing the experimental setup to provide a 12-h visible light ($\sim$ 6500lux) and 12-h darkness cycle (< 6lux), matching the timing of the light/dark cycle under which experimental flies were reared. This visible light cycle did not appreciably affect the IR imaging.

\subsection*{Temperature and humidity}

We recorded temperature and humidity within the imaging enclosure throughout the trials (Supplemental Figure 2) with a Extech RHT10 datalogger. As temperature and humidity have known effects on fly behavior \autocite{sayeed1996behavioral,Soto-Padilla2018-ff}, these data are provided with the behavioral data set so that they may be taken into account (\autoref{fig:figures9}). The environmental controls of the room in which our experiments were housed cycle on and off, leading to $\sim1^{\circ}$ C temperature fluctuations with a period of $\sim$1 hour.

\subsection*{Acquisition software}

We used a modified version of campy (\hyperlink{https://github.com/Wolfffff/campy}{github.com/Wolfffff/campy}) forked from \hyperlink{https://github.com/ksseverson57/campy}{github.com/ksseverson57/campy} which was developed by Kyle Seversson. We altered the package to suit our specific use-case, including chunking videos and adjusting the exception handling. Campy pipes frames from FLIR's Spinnaker SDK (PySpin) to FFmpeg. The flexibility of FFmpeg allows us to drastically reduce the file size of our videos by utilizing hardware-based compression. Specifically, we use Nvidia NVENC (hevc\_nvenc) paired with the segment\_time flag to produce hour-long chunks. This increased compression makes it feasible to perform high-throughput recordings of 8 flies simultaneously on a single computer. To facilitate ease of use in analysis and distribution, we merge these videos into long videos; however, because loading tens of millions of frames and instances can cause IO issues, we use hour long segments for training.

The machines used for recording were running Windows 10 with 64GB RAM, Intel(R) Core(TM) i7-8700K CPU processor, and either Nvidia Quadro RTX 4000, Quadro P2000, or GeForce RTX 2080 GPU.

\subsection*{SLEAP tracking}
After imaging, SLEAP \autocite{Pereira2022-as} was used to estimate the pose of each individual and maintain identity across videos. We used a 14 node skeleton: head, eyes (eyeL, eyeR), proboscis, thorax, abdomen, fore legs (forelegL, forelegR), mid legs (midlegL, midlegR), and the hind legs (hindlegL, hindlegR). We labeled 1930 individuals across 482 frames. 434 frames (1738 instances) were used for training, with 48 frames (192 instances) reserved for validation. We trained a U-Net based model with a receptive field size of 76 pixels ($2.\overline{6}$mm) on Nvidia A100 GPUs. The complete hyperparameter set is provided along with the model. We include some training data from recordings not included in the final data set due to early truncation but with identical frame rates and resolution. To facilitate dealing with the more than 500 million frame dataset, we use SLURM to distribute our inference across 30 Nvidia P100 GPUs at approximately 20 fps yielding approximately 600fps -- 6x speed -- tracking. After inferring locations with identity, we merged the resulting .slp files together and ran SLEAP's identity tracking script to preserve identity over time. For convenience of analysis and storage, we convert each .slp file to HDF5. Since individuals are in separate chambers, we can validate these identity tracks by the amount of time spent in each quadrant of the arena. The pipeline for sectioning, merging, and tracking can be found on the associated GitHub repository.

\subsection*{Edge detection}

While flies spend the majority of their time in the flat bottoms of the arenas, there is a small proportion of time ($\sim$5\%) when they are oriented sideways with respect to the cameras with their tarsi on the walls of the arenas. In this position the legs are often occluded and difficult to identify, leading to SLEAP tracking errors. In order to provide a flag for time points when the flies are on the edge and tracking fidelity is compromised we used the MATLAB Classification Learner App to train an SVM to identify whether flies are on or off the edge based on the all-by-all distances between tracked body coordinates (excluding the proboscis), the speed of each body coordinate, and the distance between each body coordinate and the edge of the arena. We used 2788 training points equally split between on and off edge instances, and sampled evenly across all experimental flies. Our final model accurately labeled 95\% of held out validation points (\autoref{fig:figures3}).

\subsection*{Unsupervised behavioral classification}

To identify stereotyped behaviors from body-part dynamics, we adapted the previously described MotionMapper pipeline \autocite{Berman2014-ee} for our data (\autoref{fig:figures4}). We first partially filled in missing data, interpolating all missing data for head and thorax points using Piecewise Cubic Hermite Interpolating Polynomial (PCHIP), to allow for subsequent egocentrizing. For all other nodes, we performed PCHIP interpolation with a limit of filling 5 consecutive missing values. Further, for the proboscis node, we replaced all missing values with the location of the had, representing a retracted proboscis. Further, we performed a median filter on all nodes with a window size of 5 and Gaussian filtering with standard deviation 1 and window size 5. Following this, we egocentrized the data by shifting all individuals so that the thorax is at (0, 0) and rotating each node location so that the thorax-head connection falls along the positive x-axis. After this, we calculated the Lomb-Scargle periodogram on rolling windows for each coordinate of each node. Because the Lomb-Scargle periodogram allows the utilization of unevenly sampled data and avoids the necessity of providing fully interpolated data. Further, by adjusting the window size based on our frequency of interest, we are able to capture behaviors across timescales similar to the envelope size in continuous wavelet transforms. 

We compiled a representative subsample of our data by selecting 141 fly hours evenly across flies and time of day. Because flies are dying throughout the course of the experiment our sample set is slightly skewed towards earlier days to maintain even sampling across flies. We filtered training points from this subsample of data by removing time points where the flies were on the edge. We also removed time points we classified as idle where the total amplitude of the wavelets was less than $0.5012mm^2$, a threshold we empirically determined to separate the majority of idle instances where the fly was largely motionless. From these, we sampled 36000, or the maximum number of unfiltered time points, from each fly-hour. From each of the these groups, we importance sampled 454 time points for a total of 64,014 training points.

We embedded our importance-sampled training set into two dimensions using UMAP and used this map for behavioral segmentation. We found that UMAP resulted in superior separation into unique clusters for the total training set when compared with t-SNE. We used kernel density estimation to create a 2D probability distribution of our training points. To identify distinct peaks in the density of training points we eliminated points of extreme low density and utilized adaptive thresholding on the resulting distribution. We adjusted parameters by eye to achieve distinct clusters for obviously separate peaks of density while aiming to avoid oversegmentation.

In order to assign specific discrete behaviors to each region of stereotyped power spectra we randomly selected clips from our sample set (141 fly hours) corresponding to each region. We imposed a minimum duration based on the dwell time distribution for each region to avoid very short bouts where behaviors might be difficult to identify. We identified six well-defined stereotyped behaviors (proboscis extension, fore grooming, hind grooming, wing grooming, altered locomotion, and locomotion) as well as many clusters that corresponded to idle behaviors with single-joint SLEAP tracking errors.

We then embedded our entire dataset into the same two dimensional space. Using the boundaries defined on the training set we assigned all time points to one of our six well-defined stereotyped behaviors, idle, edge (as called by our edge detector), or unstereotyped. With this method, only $\sim$15\% of our data is classified as unstereotyped behavior.

Dwell times within these behavior states can vary from single frames to many hundreds of frames. To identify a reasonable minimum bout length we fit two geometric distributions to the total dwell time histogram. We selected 5 frames ($\sim$1/20 of a second) as a minimum bout duration, as this excludes $\sim$95\% of bouts from the distribution dominated by shorter bouts, and only $\sim$14\% of bouts from the distribution of longer bouts, which we take to include legitimate behavior bouts. We forward-filled ethograms with this bout duration, assigning any bout of 4 frames or less to the previous behavior of long duration.

\section*{Data availability}

The data repository associated with this paper can be found at \hyperlink{https://doi.org/10.34770/1sab-8845}{doi.org/10.34770/1sab-8845}. For each individual, we provide a single HDF5 file that includes datasets for the tracked body parts, stereotyped behaviors, on/off edge classification, temperature and humidity data, along with experimental metadata such as start date and time and lights on and off times. Videos cropped to contain individual flies are also provided. The original uncropped videos and the full postural tracking data, as .slp files with prediction scores for each body part of each individual, are available upon request.

\section*{Code availability}

The source code for the data analysis is publicly available. The code can be found on GitHub (\hyperlink{https://github.com/shaevitz-lab/long-timescale-analysis}{github.com/shaevitz-lab/long-timescale-analysis}). The repository includes the scripts used in this paper along with other pragmatic tools and examples.

The modified version of MotionMapperPy \autocite{Berman2014-ee} we use can be found at \hyperlink{https://github.com/Wolfffff/motionmapperpy}{https://github.com/Wolfffff/motionmapperpy} and included as a git submodule in the primary repository.

\section*{Acknowledgements}

The authors acknowledge the Aspen Center for Physics where this work was first conceptualized, Gordon Berman, Ugne Klibaite, and Greg Stephens for inspirational discussion, and Diogo Melo for insightful comments on how to speed up our processing pipeline. This work was supported in part by the NSF through the Center for the Physics of Biological Function (PHY-1734030). SWW is supported by the NSF Graduate Research Fellowship Program (DGE-2039656). GCM-S is supported by the Paul F. Glenn Laboratories For Aging Research at Princeton University. JFA is funded by grants from the NIH: National Institute of Environmental Health Sciences (R01-ES029929) and National Institute of General Medical Sciences (R35GM124881). We also acknowledge that the work reported in this paper was substantially performed using the Princeton Research Computing resources at Princeton University, which is a consortium of groups led by the Princeton Institute for Computational Science and Engineering (PICSciE) and the Office of Information Technology's Research Computing group.

\section*{Author contributions statement}

% Must include all authors, identified by initials, for example:

Conceptualization, JWS and SWW; Initial methodology, SWW; Methodology development, SWW and GCM-S; Investigation, SWW and GCM-S; Formal analysis, SWW and GCM-S; Resources, JFA and JWS; 
Writing-Original draft, SWW and GCM-S; Writing-Review \& Editing, SWW, GCM-S, JFA and JWS; Funding Acquisition, SWW, JFA, and JWS.

\section*{Competing interests}
The authors declare no competing interests.

\printbibliography
%  CONFIGURE NEW SINGLE-PAGE FORMAT 

\onecolumn % go back to one column
\fancyhead{} % make sure we get no headers
\renewcommand{\floatpagefraction}{0.1}
\lfoot[\bSupInf]{\dAuthor}
\rfoot[\dAuthor]{\cSupInf}
\newpage

\captionsetup*{format=largeformat} % make figure legend slightly larger than in the paper
\setcounter{figure}{0} % reset figure counter for Supp. Figures
\setcounter{equation}{0} % reset equation counter for Supp. Equations
\makeatletter
\renewcommand{\thefigure}{S\@arabic\c@figure} % make Figure legend start with Figure S
\makeatother
\def\theequation{S\arabic{equation}}

%  MAIN TEXT 

\newpage
\section*{Supplementary Information}

\section*{Tables}

\begin{table}[ht]
\resizebox{\textwidth}{!}{
\begin{tabular}{|c|c|c|c|c|}
\hline
\textbf{Date} & \textbf{Experiment} & \textbf{Camera} & \textbf{Start Time (UTC)}  & \textbf{File}     \\ \hline
2/17/2022     & exp1                & Camera 1        & 2022/02/17 17:37:18 & 20220217-lts-cam1 \\ \hline
2/17/2022     & exp1                & Camera 2        & 2022/02/17 17:37:18 & 20220217-lts-cam2 \\ \hline
2/17/2022     & exp1                & Camera 3        & 2022/02/17 17:46:39 & 20220217-lts-cam3 \\ \hline
2/17/2022     & exp1                & Camera 4        & 2022/02/17 17:46:39 & 20220217-lts-cam4 \\ \hline
3/13/2022     & exp2                & Camera 3        & 2022/03/13 02:14:15 & 20220313-lts-cam3 \\ \hline
3/13/2022     & exp2                & Camera 4        & 2022/03/13 02:14:15 & 20220313-lts-cam4 \\ \hline
3/26/2022     & exp3                & Camera 3        & 2022/03/26 19:50:45 & 20220326-lts-cam3 \\ \hline
3/26/2022     & exp3                & Camera 4        & 2022/03/26 19:50:45 & 20220326-lts-cam4 \\ \hline
4/18/2022     & exp4                & Camera 1        & 2022/04/18 19:06:00 & 20220418-lts-cam1 \\ \hline
4/18/2022     & exp4                & Camera 2        & 2022/04/18 19:06:00 & 20220418-lts-cam2 \\ \hline
4/18/2022     & exp4                & Camera 3        & 2022/04/18 19:07:28 & 20220418-lts-cam3 \\ \hline
4/18/2022     & exp4                & Camera 4        & 2022/04/18 19:07:28 & 20220418-lts-cam4 \\ \hline
\end{tabular}
}
\caption{Metadata table outlining the data collected. Each row represents a single camera recording on a single day covering 4 wells. Here, we show the composition of experiments starting from mid day to evening. The provided File column corresponds to the H5 file containing tracks of each recording. Complete metadata at the per-fly level is provided in the associated data repository.}
\label{tab:metadata}
\end{table}
\pagebreak
\section*{Figures}

\begin{figure*}[!ht]
	\centering
	\includegraphics[width=\linewidth]{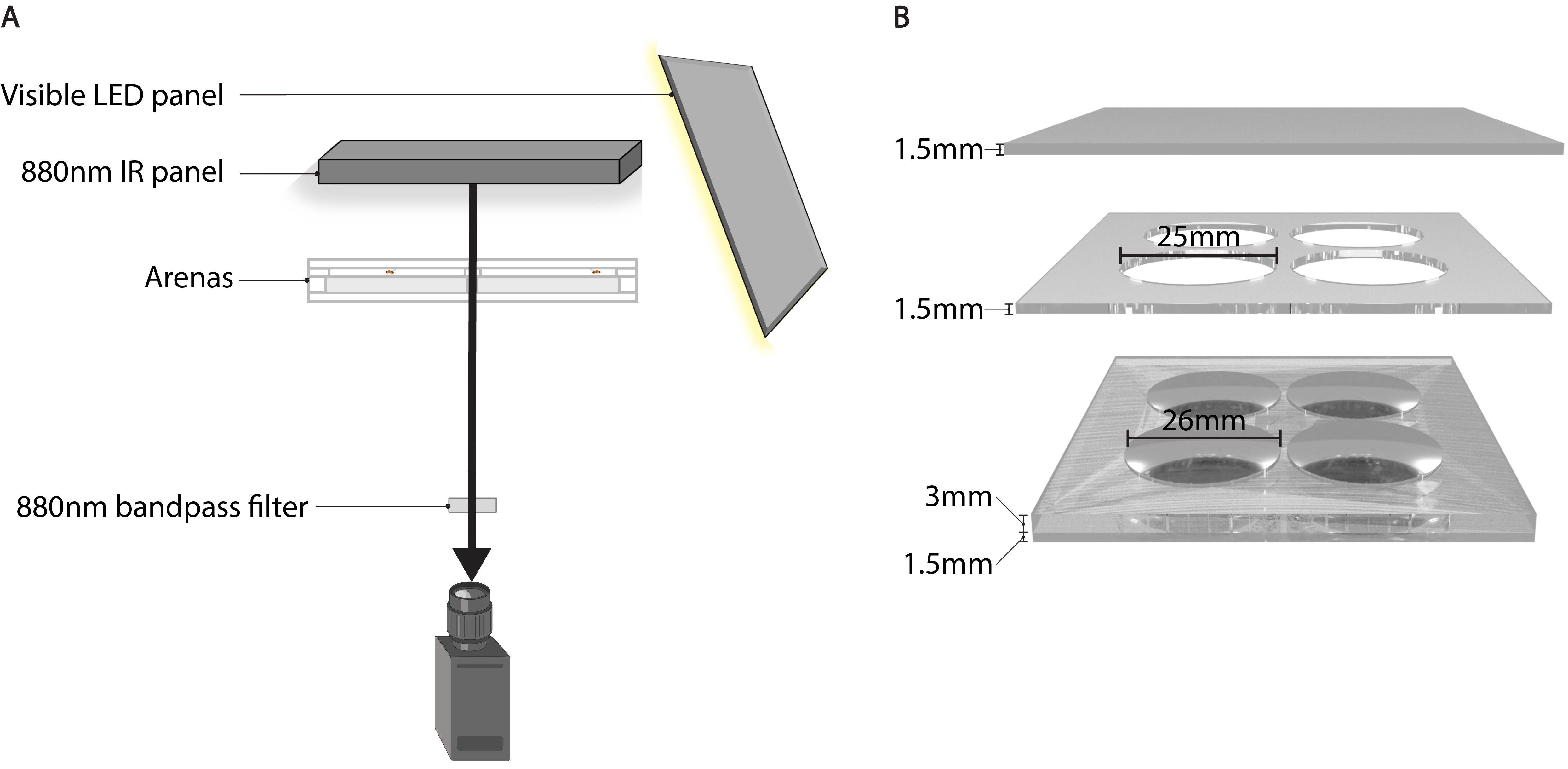}
	\caption{\textbf{A} Schematic of imaging setup. Experimental arenas were illuminated from above with 880nm LED pads to permit constant recording and two visible LED panels (only one shown) on a 12-hour light/dark cycle. Flies were recorded from below using 880nm bandpass filters on each camera to ensure uniformity across visible light changes. \textbf{B} Experimental arena schematic. Arenas were constructed from layers of transparent laser-cut acrylic, with a 3mm deep pad of sucrose-agarose media beneath a 1.5mm deep chamber enclosed by a solid layer of acrylic. Arena layers were held together by lab tape, which prevented escape while also permitting airflow.}
	\label{fig:figures1}
\end{figure*}

\begin{figure*}[ht]
  \centering
  \includegraphics[width=\linewidth]{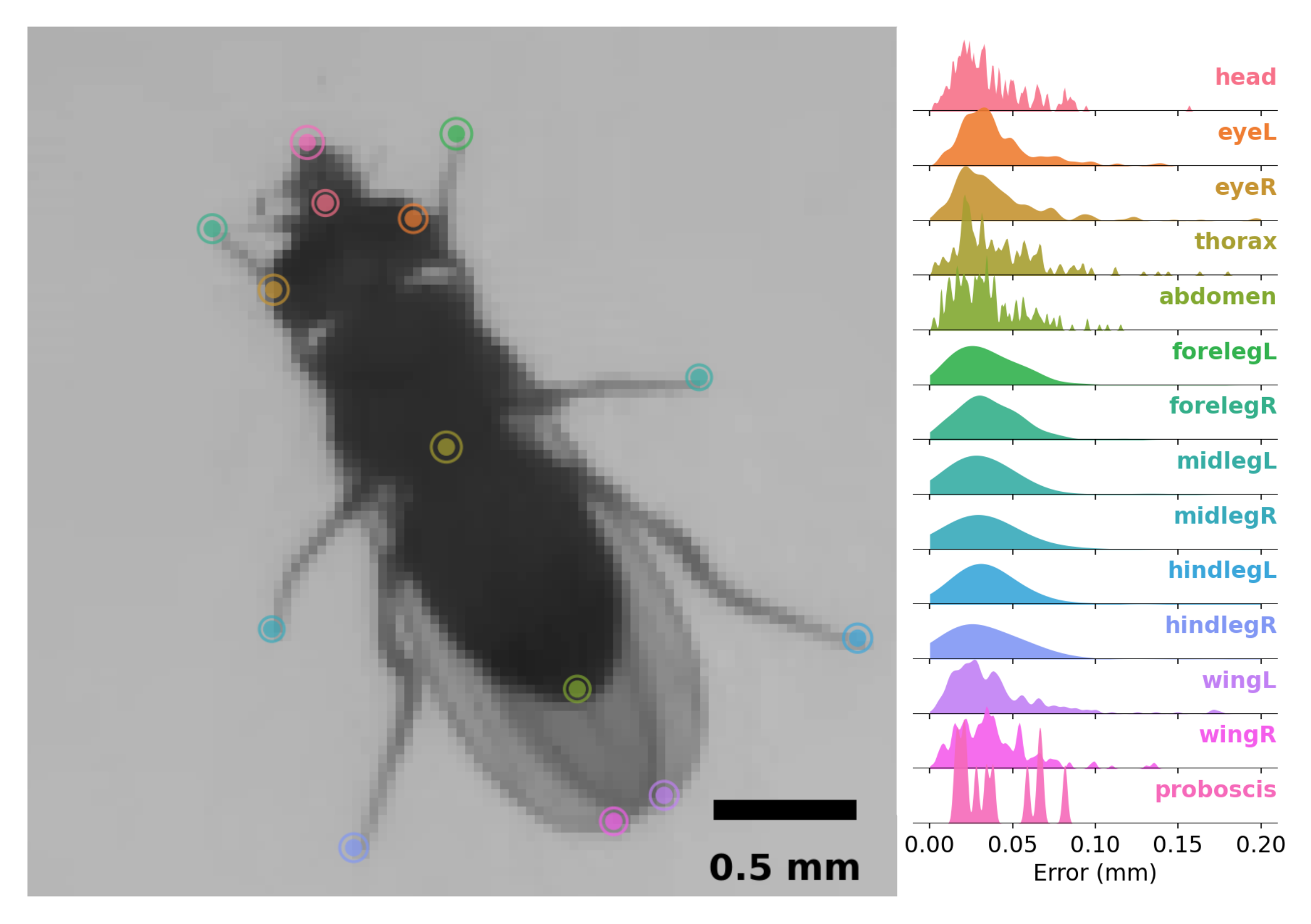}
  \caption{Prediction error plot. The average error distance is approximately 2.22px, corresponding to 78.5µm. Our model's mean average precision (mAP) is 0.70, and the error distance 95th percentile is 4.14px. More metrics and complete models are available in the main dataset.}
  \label{fig:figures2}
\end{figure*}

\begin{figure*}[ht]
  \centering
  \includegraphics[width=\linewidth]{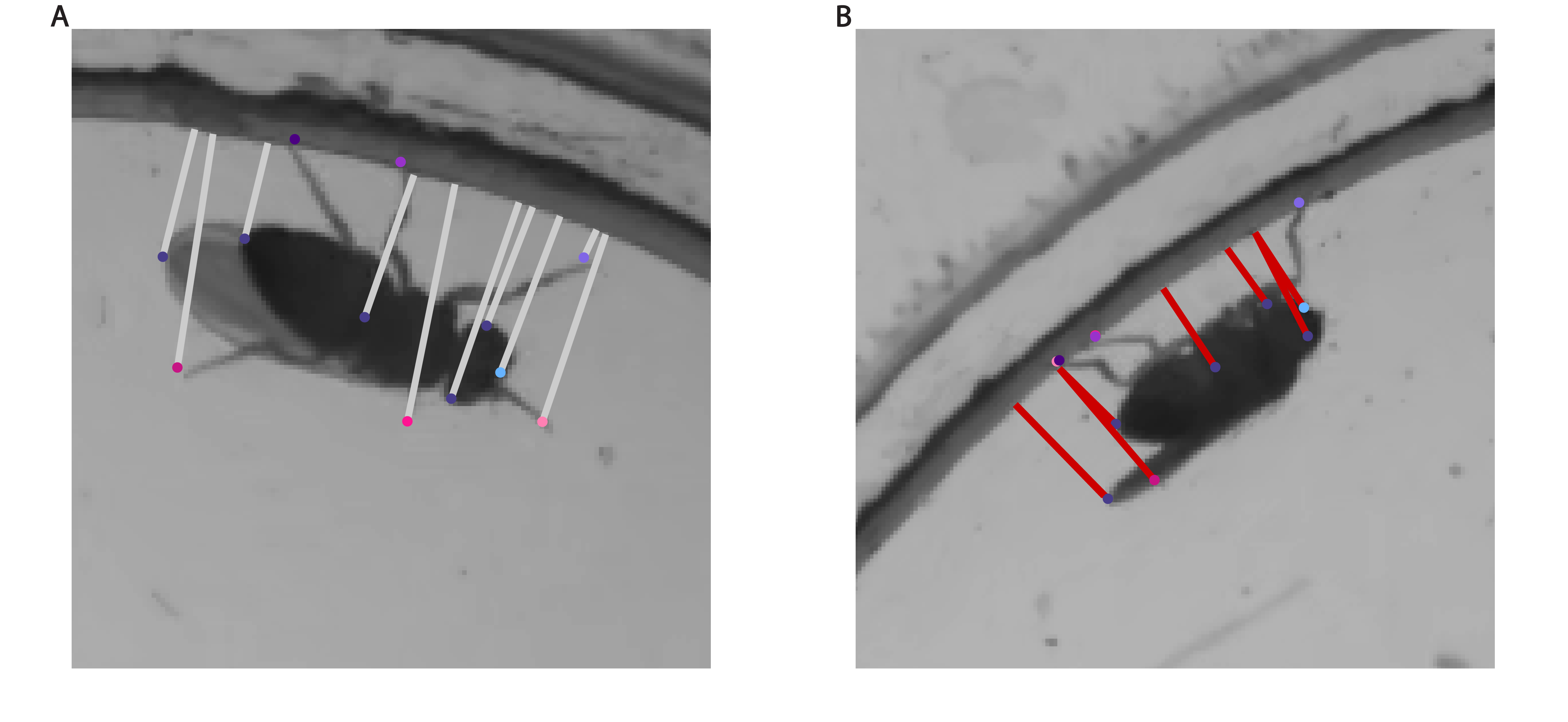}
  \caption[Illustration of edge detection method.]{Illustration of edge detection method. An SVM classifier uses the all-by-all distances between all body part coordinates, except for the proboscis, the speed of each body part, and the distance for each body part from the edge to classify time points as 'off edge' (example shown on left) or 'on edge' (example shown on right). Points directly on the edge, such as some of the tarsi in these images, have an edge distance of 0, which naturally cannot be shown.}
  \label{fig:figures3}
\end{figure*}

\begin{figure*}[ht]
  \centering
  \includegraphics[width=\linewidth]{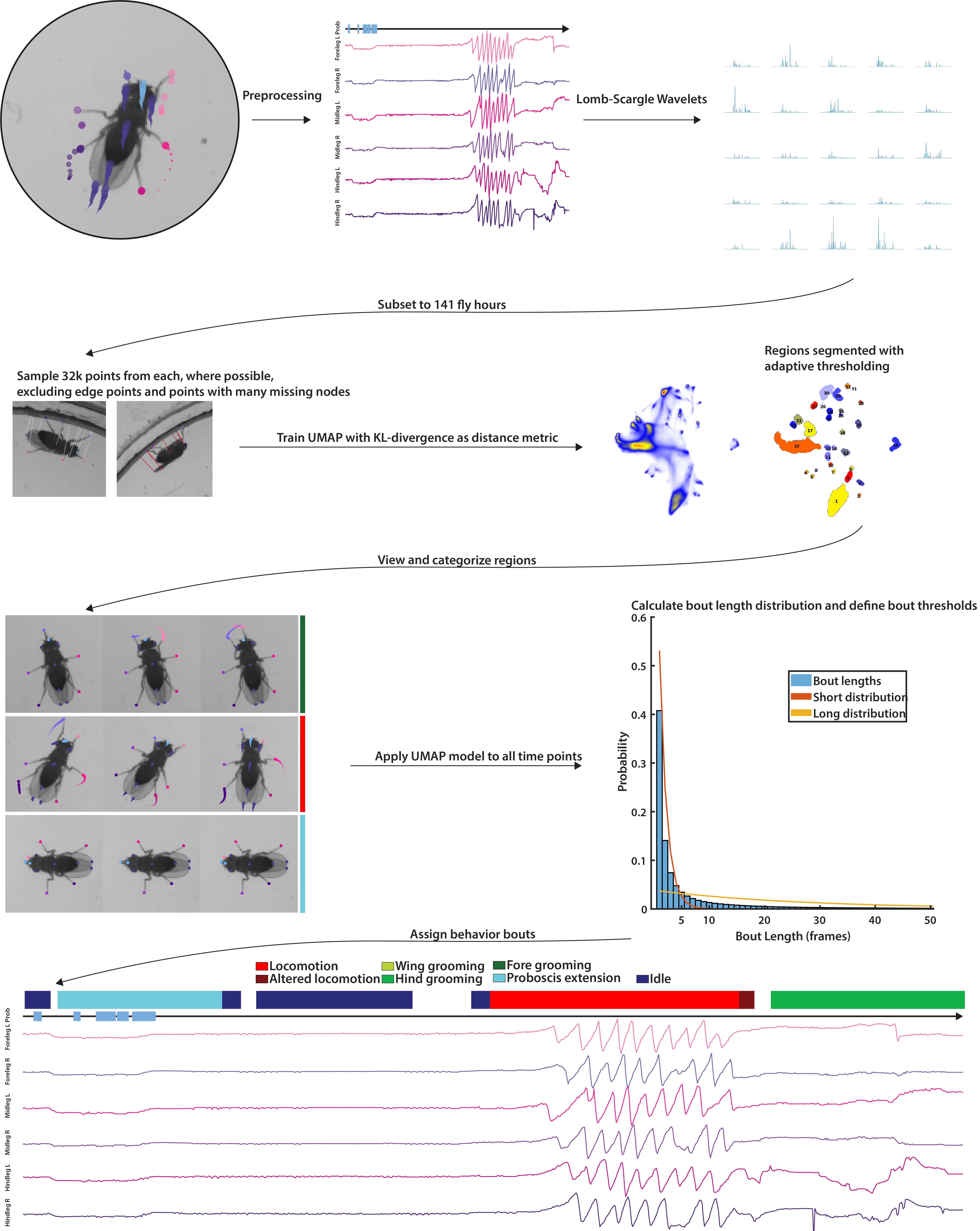}
  \caption[Schematic of behavioral classification pipeline. ]{Schematic of behavioral classification pipeline. The behavioral pipeline shows the flow of data from pose estimation from SLEAP through behavioral bout assignment.}
  \label{fig:figures4}
\end{figure*}

\begin{figure}[!ht]
	\centering
	\includegraphics[width=\linewidth]{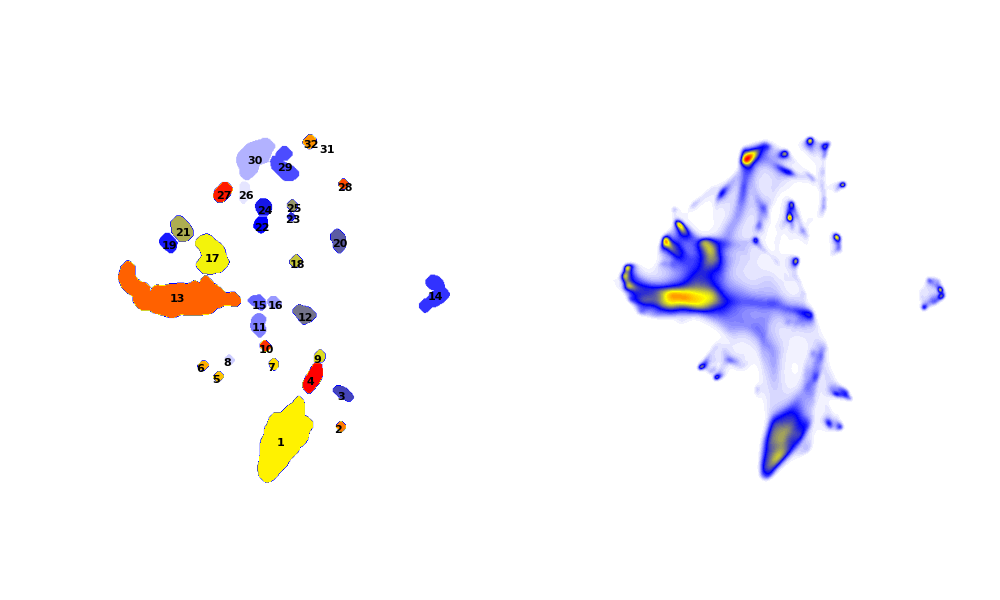}
	\caption[Plots showing the density map of 2D embedding values from UMAP along with the region assignments.]{Plot showing the density map of 2D embedding values from UMAP along with the region assignments. Regions 2, 3, 4, 5, 9, 12, 15, 16, 18, 20, 23, 25, 28, 31, and 32 were assigned to idle behavior, 14 was assigned to proboscis extension. Regions 1, 6, 7, 8, 10, and 11 were assigned to foreleg grooming. Regions 26, 27, and 30 were assigned to hind grooming. Regions 22, 24, and 29 were assigned to wing grooming behavior. Regions 17, 19, and 21 were assigned to altered locomotion. Finally, region 13 was assigned to locomotion.}
	\label{fig:figures5}
\end{figure}

\begin{figure*}[ht]
  \centering
  \includegraphics[width=\linewidth]{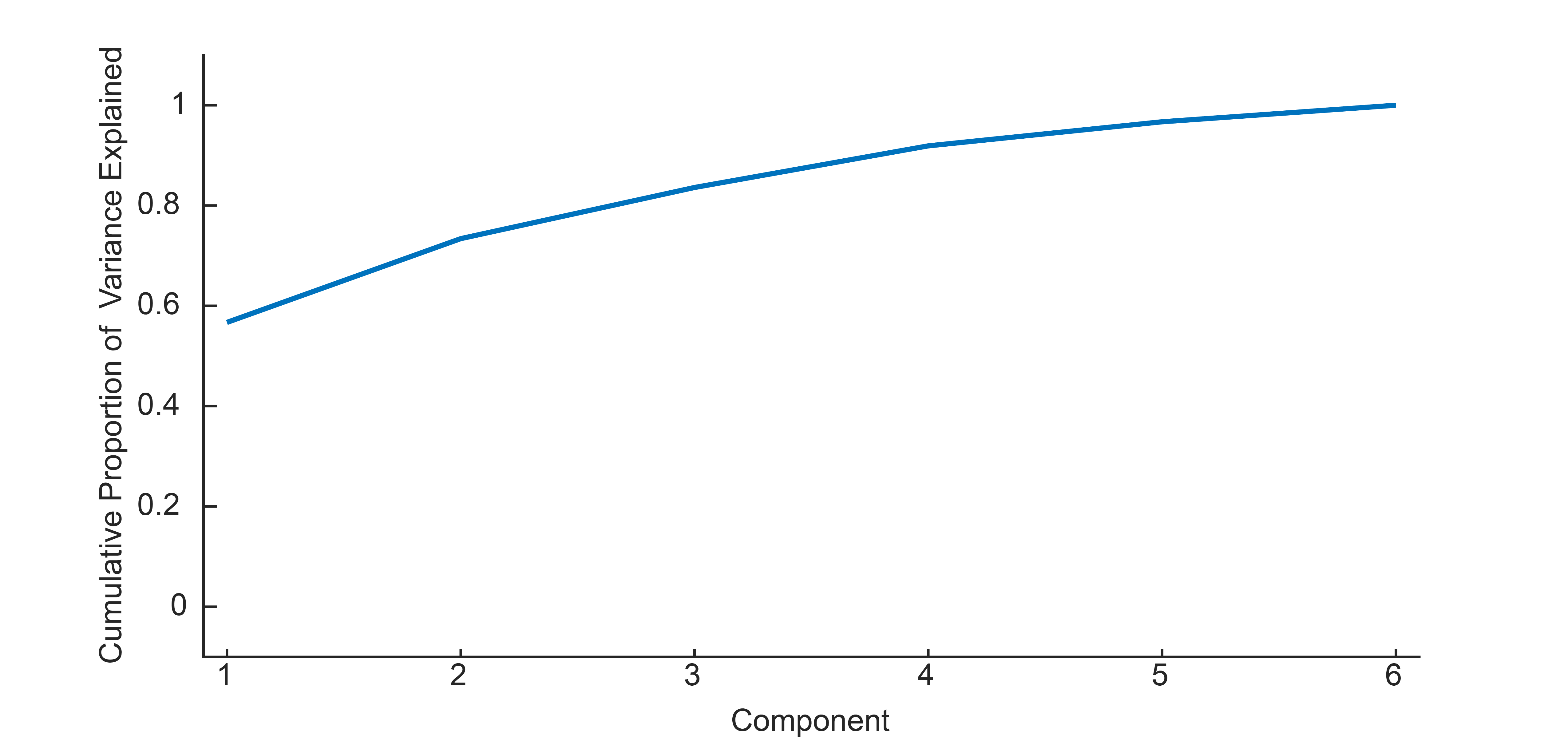}
  \caption[Line plot showing the cumulative variance explained by PCs of behavioral components complete data set.]{Line plot showing the cumulative variance explained by PCs of behavioral components on the complete data set.}
  \label{fig:figures6}
\end{figure*}

\begin{figure*}[ht]
  \centering
  \includegraphics[width=\linewidth]{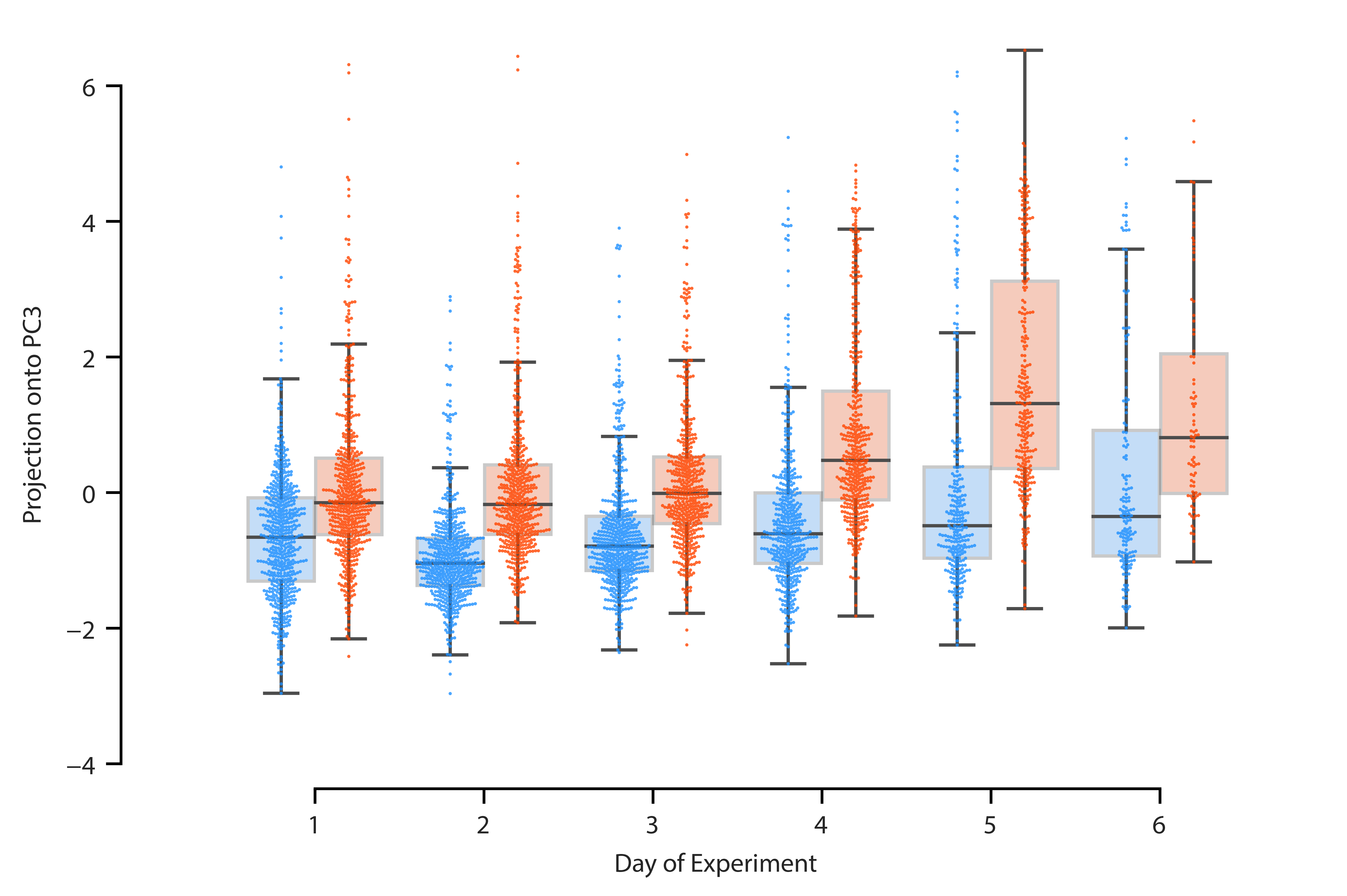}
  \caption[Box plots showing PC3 from behavioral components PCA.]{Box plots showing PC3 of behavioral components PCA. Experimental group 1 (experiments 1 and 2) is shown in blue and experimental group 2 (experiments 3 and 4) are shown in red. PC3 separates our experimental groups and this separation becomes more significant farther into the experiments. }
  \label{fig:figures7}
\end{figure*}

\begin{figure*}[ht]
  \centering
  \includegraphics[width=\linewidth]{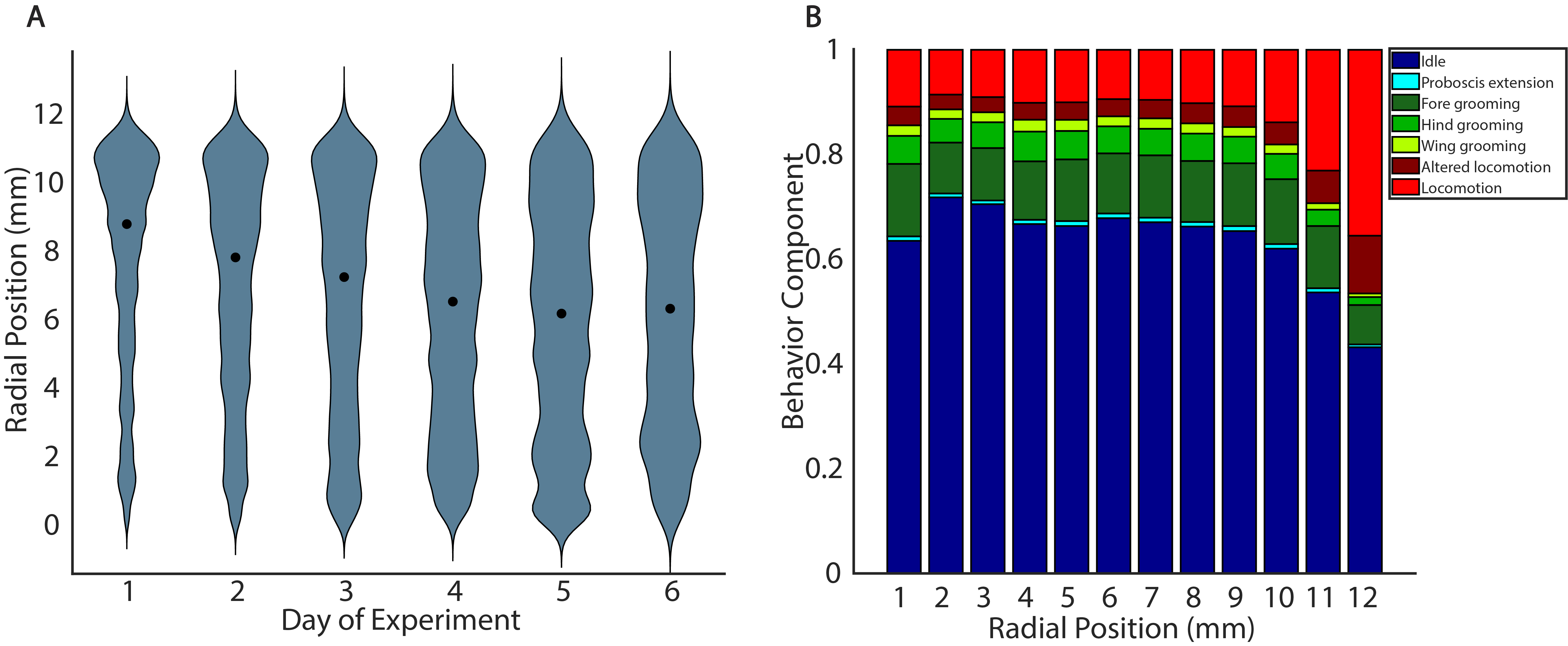}
  \caption[Behavioral characteristics by radial position and radial position distributions by day of experiment.]{Behavioral characteristics by radial position and radial position distributions by day of experiment. \textbf{A} Violin plot showing the distributions of radial position by day of experiment. \textbf{B} Barplot of behavioral components by radial position.}
  \label{fig:figures8}
\end{figure*}

\begin{figure*}[ht]
  \centering
  \includegraphics[width=\linewidth]{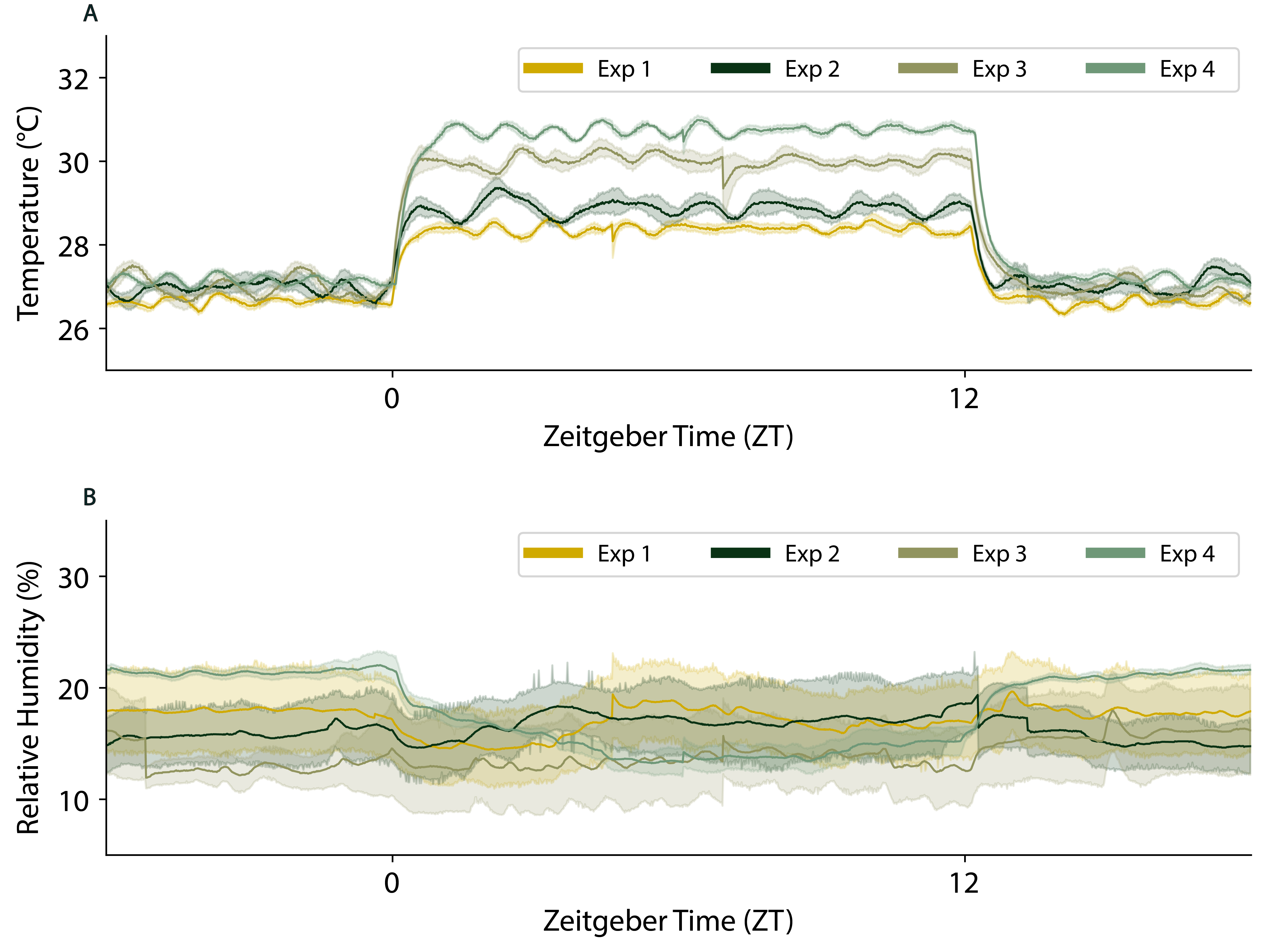}
  \caption[Line plots showing temperature and humidity measured throughout the experiments.]{Line plots showing temperature and humidity measured throughout the experiments. All measurements were taken a 1 minute intervals and have been trimmed to only include points where at least one fly in the experiment is alive. \textbf{A} Line plot showing the variation in temperature across experiments and zeitgeber time. The shaded region is the standard error. \textbf{B} Line plot showing the variation in humidity across experiments and zeitgeber time. The shaded region is the standard error.}
  \label{fig:figures9}
\end{figure*}
\end{document}